\documentclass[fleqn,usenatbib]{mnras}

\usepackage{newtxtext,newtxmath}

\usepackage[T1]{fontenc}
\usepackage{xcolor}
\usepackage{ulem} 

\DeclareRobustCommand{\VAN}[3]{#2}
\let\VANthebibliography\thebibliography
\def\thebibliography{\DeclareRobustCommand{\VAN}[3]{##3}\VANthebibliography}

\defcitealias{Yu2017}{Y17}
\defcitealias{Dalal2008}{D08}
\defcitealias{Lazeyras2021}{L21}

\newcommand{\lfs}{\lambda_{\mathrm{fs}}}
\newcommand{\mnu}{\Sigma{M_{\nu}}}
\newcommand{\nuplus}{M_{\nu}^{+++}}
\newcommand{\nuzero}{{\nu}0.0}
\newcommand{\nufour}{{\nu}0.4}
\newcommand{\ev}{\mathrm{eV}}
\newcommand{\hmsol}{h^{-1}M_\odot}
\newcommand{\hmpc}{h^{-1}\mathrm{Mpc}}

\usepackage{graphicx}	
\usepackage{amsmath}	
\usepackage{comment}

\title[Impact of neutrinos on halo assembly bias]{Elucidating the impact of massive neutrinos on halo assembly bias}

\author[Y. Song \& Y. Zu]{
Yunjia Song,$^{1}$ and
Ying Zu$^{1,2,3,4}$\thanks{E-mail: yingzu@sjtu.edu.cn}
\\
$^{1}$Department of Astronomy, School of Physics and Astronomy, Shanghai Jiao Tong
University, Shanghai 200240, China\\
$^{2}$Tsung-Dao Lee Institute, Shanghai Jiao Tong University, Shanghai
200240, China\\
$^{3}$Shanghai Key Laboratory for Particle Physics and Cosmology, Shanghai Jiao Tong
University, Shanghai 200240, China\\
$^{4}$Key Laboratory for Particle Physics, Astrophysics and Cosmology,
Ministry of Education, Shanghai Jiao Tong University, Shanghai 200240
}

\date{Accepted XXX. Received YYY; in original form ZZZ}

\pubyear{2023}

\begin{document}
\label{firstpage}
\pagerange{\pageref{firstpage}--\pageref{lastpage}}
\maketitle

\begin{abstract}
    Massive neutrinos have non-negligible impact on the formation of
    large-scale structures.  We investigate the impact of massive neutrinos
    on the halo assembly bias effect, measured by the relative halo bias
    $\hat{b}$ as a function of the curvature of the initial density peak
    $\hat{s}$, neutrino excess $\epsilon_\nu$, or halo concentration
    $\hat{c}$, using a large suite of $\Sigma M_\nu{=}0.0$ eV and $0.4$ eV
    simulations with the same initial conditions. By tracing dark matter
    haloes back to their initial density peaks, we construct a catalogue of
    halo ``twins'' that collapsed from the same peaks but evolved
    separately with and without massive neutrinos, thereby isolating any
    effect of neutrinos on halo formation. We detect a
    $2\%$ weakening of the halo assembly bias as measured by
    $\hat{b}(\epsilon_\nu)$ in the presence of massive neutrinos.
    As there exists a significant correlation between $\hat{s}$
    and $\epsilon_\nu$~($r_{cc}{=}0.319$),
    the impact of neutrinos persists
    at a reduced level~($0.1\%$)
    in the halo assembly bias measured by $\hat{b}(\hat{s})$.
    However,
    we do not detect any neutrino-induced impact on
    $\hat{b}(\hat{c})$, consistent with earlier studies and the lack of
    correlation between $\hat{c}$ and $\epsilon_\nu$~($r_{cc}{=}0.087$).
    We also discover an analogous assembly bias effect for the ``neutrino
    haloes'', whose concentrations are anti-correlated with the large-scale
    clustering of neutrinos.
\end{abstract}




\section{Introduction}
\label{sec:intro}

Massive neutrinos play a key role in the formation of large-scale
structures~(LSS) of the Universe~\citep{Lesgourgues2006,Wong2011}.  Their
lack of clustering on scales below the free-streaming length induces a
scale-dependent growth of the LSS, which enables stringent constraints of
the sum of neutrino masses $\mnu$ with cosmological
probes~\citep{Cuesta2016, Vagnozzi2017, Brinchmann2019, Dvorkin2019,
Giusarma2016, Planck2020, Alam2017, Mikhail2020, Palanque-Delabrouille2020,
DiValentino2021}. Another important effect caused by the massive neutrinos
is the differential growth of dark matter haloes between environments of
different neutrino-to-dark matter density ratios~\citep{Yu2017}. Through
this effect, the presence of massive neutrinos may cause an impact on the
so-called ``halo assembly bias''~\citep[hereafter referred to as
HAB;][]{Gao2007}, potentially opening a new avenue for measuring $\mnu$
with LSS observations. However, such impact can only be weak at best, as it
has so far evaded detections in the simulations~\citep{Lazeyras2021}.  In
this paper, we focus on the ``twin'' haloes collapsed from the same
initial density peaks but evolved separately with and without massive
neutrinos, aiming to elucidate the impact of massive neutrinos on the
assembly bias of massive haloes.

HAB has been robustly predicted by the hierarchical structure formation of
$\Lambda$+cold dark matter~(CDM)
simulations~\citep{Sheth2004,Gao2005,Wechsler2006,Harker2006,Jing2007,Li2008}.
It has increasingly been referred to as the ``secondary
bias''~\citep{Salcedo2018,Mao2018,Chue2018,Montero-Dorta2020,Contreras2021,Lin2022,Lazeyras2023,Wang2023}
--- the large-scale bias of haloes depends not only on halo mass, but also
on other halo properties such as concentration, formation time, and spin.
At any given halo mass $M_h$, the strength of HAB is often characterised by
the {\it relative} halo bias $\hat{b}$ as a function of some ``indicator''
property $x$,
\begin{equation}
    \hat{b}(x | M_h) = {b(x | M_h)}\,/\,{\langle b | M_h\rangle},
\end{equation}
where $b(x | M_h)$ is the bias of haloes with $x$ and $\langle b |
M_h\rangle$ is the overall bias at $M_h$. \citet{Contreras2021} found that
HAB is independent of cosmological parameters in the $\Lambda$CDM, at least
for the high-mass haloes that we focus on in this paper. Such an
independence on cosmology could be useful for measuring $\mnu$, if HAB
turns out to be sensitive to massive neutrinos.

Observationally, HAB at the high mass end has been at the brink of
detection recently~\citep{Yang2006, Lin2016, Miyatake2016, Zu2017, Lin2022,
Zu2021, Sunayama2022, Zu2022}. Using cluster weak lensing, \citet{Zu2021}
found that the stellar mass of the central galaxies is a good observational
proxy for concentration. Combining with the $\hat{b}$ measurements from
cluster-galaxy cross-correlations, \citet{Zu2022} detected a possible
evidence of the cluster assembly bias, thereby demonstrating a viable path
to accurate HAB measurements using future cluster surveys. In particular,
$\Lambda$CDM simulations predict that the cluster-mass haloes with higher
concentrations have a lower bias than their low-concentration counterparts.
\citet[][hereafter referred to as~\citetalias{Dalal2008}]{Dalal2008}
explained this phenomenon using the peak formalism~\citep{Kaiser1984,
Bardeen1986, ShethTormen1999}, which predicts that the curvature of initial
density peaks is correlated with the large-scale overdensity at fixed peak
height. Using simulations, \citetalias{Dalal2008} demonstrated that the
rare peaks with steep curvatures are more likely to collapse into clusters
with high concentrations in low-density environments, and vice versa for
those with shallow curvatures. In this paper, we follow
\citetalias{Dalal2008} to trace haloes back to their initial density peaks
and employ their peak curvature as a key indicator of HAB.

Although the presence of massive neutrinos delays the collapse of initial
density peaks into haloes, the collapse time remains accurately predicted
by the excursion set theory~\citep{Press1974,Bond1991} if the variance of
the CDM~(assuming baryons follow the CDM) rather than the total mass field
is considered~\citep{Ichiki2012}. Therefore, it is perhaps unsurprising
that \citet{Lazeyras2021} did not find any impact of neutrinos on HAB.
However, \citet{Yu2017} discovered that the mass of haloes in neutrino-rich
environments tends to be systematically higher than those in neutrino-poor
regions, probably due to the higher capturing rate of the slow-moving
neutrinos in neutrino-rich environments. If halo assembly histories are
coherently modulated by neutrinos across different environments, it is
reasonable to expect such a modulation to leave some discernible imprint on
the HAB in simulations.  To isolate the impact of neutrinos, we construct a
catalogue of ``twin'' haloes that originated from the same initial density
peaks but evolved separately in massless vs. massive neutrino
cosmologies.  Therefore, any difference of HAB between the two twin samples
would provide a smoking gun evidence of the impact of neutrinos on HAB.

This paper is organized as follows. We describe the construction of our
peak-based twin halo catalogue in Section~\ref{sec:data} and present the
HAB comparison using different indicators in Section~\ref{sec:result}. We
summarize our results and look to the future in
Section~\ref{sec:conclusion}. Throughout this paper, we adopt a spherical
overdensity-based halo definition so that the inner overdensity within halo
radius~$r_{200m}$ is 200 times the mean density of the Universe, and the
halo mass~$M_h$ is defined as the total mass enclosed by $r_{200m}$.

\section{Peak-based Twin Halo Catalogue}
\label{sec:data}

We compare the HAB effects with and without massive neutrinos using the
\texttt{Quijote} simulations~\citep{Francisco2020}, the same suite used by
\citet{Lazeyras2021}.  Among the vast \texttt{Quijote} ensemble, we select
one ``Fiducial'' set of 40 $\Lambda$CDM simulations with $\mnu{=}0.0\,\ev$,
and another ``$\nuplus$'' set of 40 massive neutrino simulations with
$\mnu{=}0.4\,\ev$, hereafter referred to as the $\nuzero$ and $\nufour$
simulations, respectively. The $\nufour$ simulations bear the highest
$\mnu$ within the \texttt{Quijote} suite, so that any possible
neutrino-induced HAB would exhibit the strongest signal. Both the CDM and
neutrinos are particle-based, with $512^3$ CDM particles in a comoving
volume of $1\,h^{-3}\mathrm{Gpc}^3$ of each simulation and an additional
$512^3$ neutrino particles for each $\nufour$ run.
The two sets of simulations have the same cosmological parameters, with
$\Omega_m{=}0.3175$, $\Omega_b{=}0.049$, $h{=}0.6711$, $n_s{=}0.9624$,
$\sigma_8=0.834$, and $\omega{=}{-}1$. More important, identical CDM
initial conditions~(with the same phases) at $z{=}127$ are shared by each
of the 40 pair of $\nuzero$ and $\nufour$ simulations, so that the haloes
found at $z{=}0$ can always be mapped back to the initial density peaks in
the $z{=}127$ snapshots. We identify dark matter haloes in the $z{=}0$
outputs of the 40 pairs of simulations using
\texttt{ROCKSTAR}~\citep{Behroozi2013a}. We calculate halo concentration
$c$ using the maximum circular velocity method~\citep{Prada2012}, and use
the {\it relative} concentration $\hat{c}$, defined as
\begin{equation}
    \hat{c} \equiv \frac{c - \langle c | M_h \rangle}{\sigma_c(M_h)},
    \label{eqn:chat}
\end{equation}
as an indicator of HAB, where $\langle c | M_h\rangle$ and $\sigma_c(M_h)$
are the average concentration-mass relation and its $1{-}\sigma$ scatter in
either the $\nuzero$ or $\nufour$ simulations.

\begin{figure}
	\includegraphics[width=0.96\columnwidth]{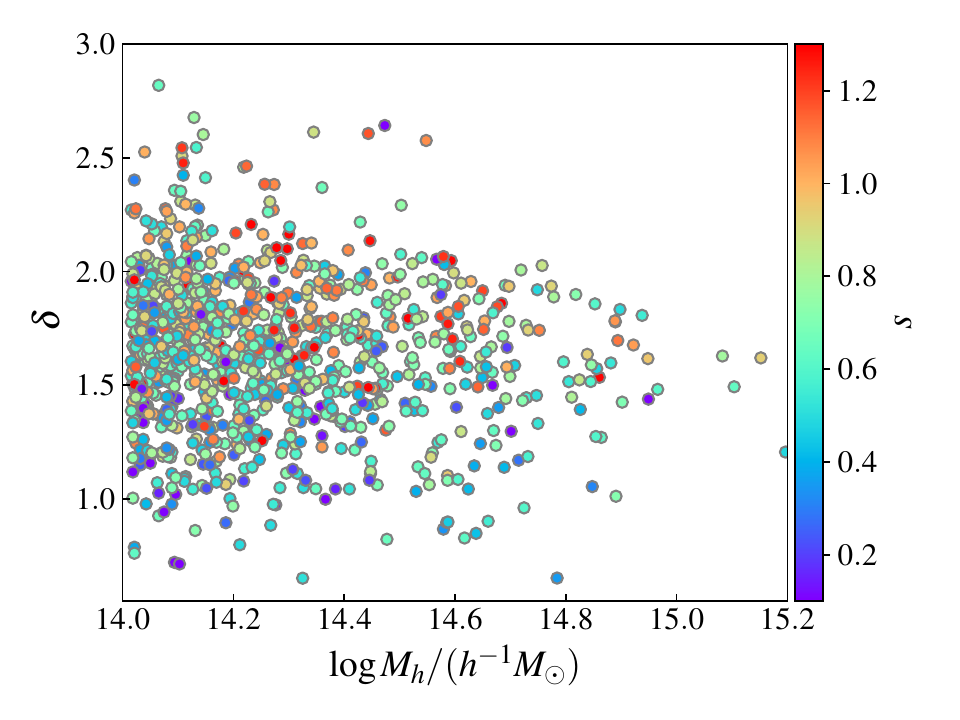}
    \caption{Distribution of 1000 haloes randomly drawn from the twin halo
    catalogue on the CDM Lagrangian overdensity $\delta$~(linearly
    extrapolated to $z{=}0$) vs. halo mass $M_h$ plane, colour-coded by
    their peak curvature $s$ according to the colourbar on the right.}
    \label{fig:curvature_colormap}
\end{figure}

\begin{figure*}
	\includegraphics[width=0.9\linewidth]{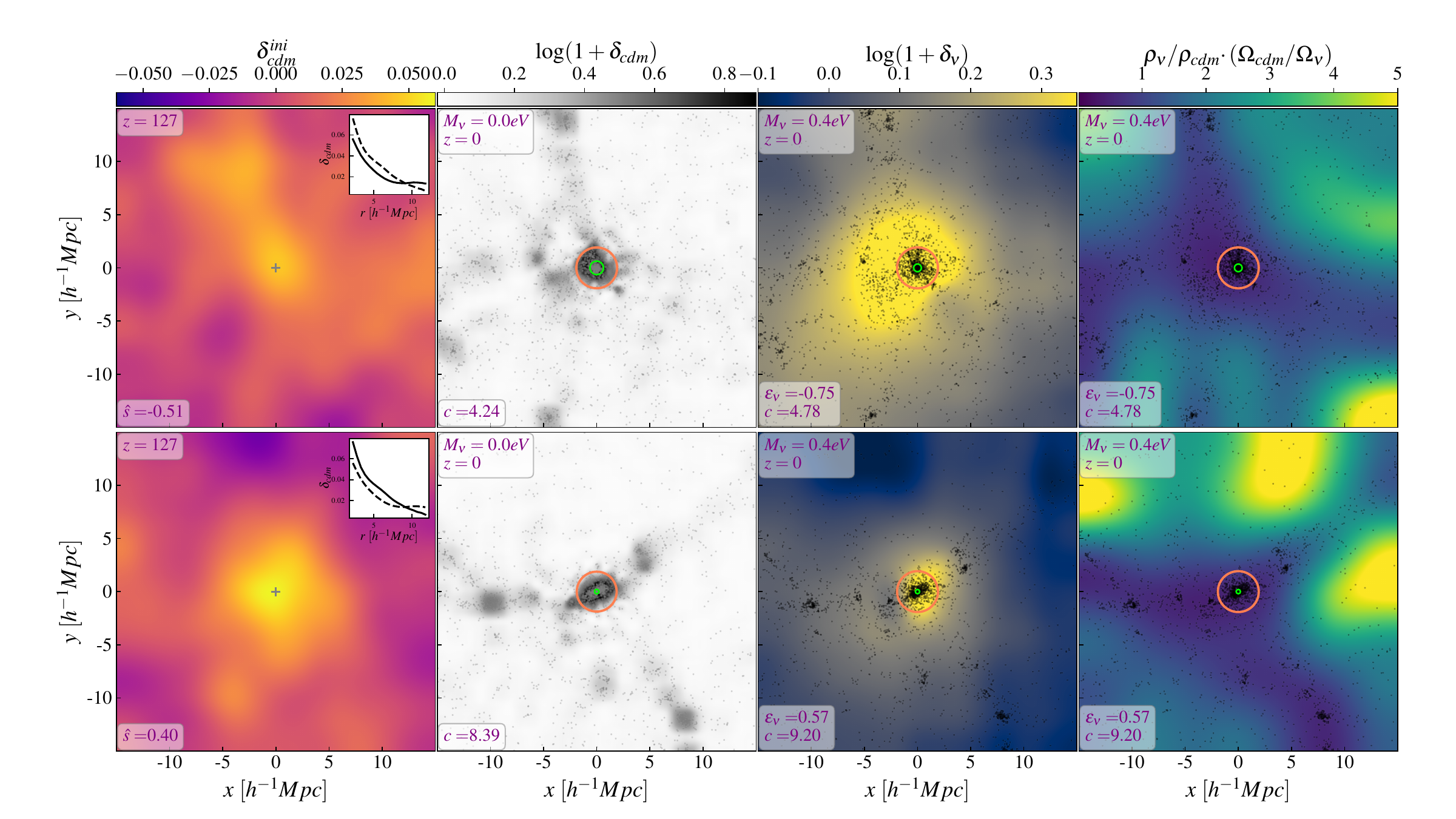}
    \caption{Distributions of the initial CDM overdensity
    $\delta^{\mathrm{ini}}_{\mathrm{cdm}}$~(leftmost), current-day CDM
    density~(2nd from left), neutrino density~(2nd from
    right), and neutrino-to-CDM density ratio
    $\rho_\nu/\rho_\mathrm{cdm}$~(normalised by
    $\Omega_\nu/\Omega_{\mathrm{cdm}}$; rightmost) around two pairs of twin
    haloes with the same peak height, one collapsed from a
    shallow~($\hat{s}{=}{-}0.51$; top) density peak and the other
    steep~($\hat{s}{=}0.40$; bottom). Each panel shows a slice of dimension
    $30\,\hmpc{\times}30\,\hmpc{\times}12\,\hmpc$ centred on the peak/halo,
    with the key halo/peak properties annotated in the bottom left corner
    and the simulation/redshift information in the top left. The maps are
    colour-coded according to the
    respective colourbars on top.  The inset panel in each of the leftmost
    panels shows the radial profiles of the CDM overdensity around the two
    peaks, with the solid curve corresponding to the
    $\delta_{\mathrm{cdm}}^{\mathrm{ini}}$ distribution shown in the main
    panel.  The inner lime and outer orange circles on the rest of the
    panels denote the scale radii and $r_{200m}$ of the haloes,
    respectively. } \label{fig:colormap}
\end{figure*}

To construct a high-purity twin halo catalogue from the paired simulations,
we proceed step by step as follows. For each halo with
$M_h{\geq}10^{14}\,\hmsol$ at $z{=}0$ in the $\nuzero$ simulations, we
collect all its CDM particles within $r_{200m}$, retrieve their initial
positions at $z{=}127$, and compute the barycentre of those particles in
the initial density field. We then identify the highest peak within a
search radius of $r_{\mathrm{pk}}=200^{1/3}r_{200m}$ about the barycentre
as the true density peak that collapsed into that halo.  The peak
identification is performed over the smoothed density field derived with
\texttt{Nbodykit}~\citep{Hand2018}. Since the exact density peak also
exists in the initial condition of the paired $\nufour$ simulation, we
collect all the CDM particles within $r_{\mathrm{pk}}$ about the peak in
the initial condition, and identify its matched counterpart as the halo
that inherited the highest fraction of those particles at $z{=}0$ in the
$\nufour$ simulation. To make sure the matches are unique, we repeat the
above procedures but start from the haloes in the $\nufour$ simulations. We
only keep the $97\%$ twins that are reproduced during the reversed search
in our final twin catalogue, leaving $3458747$ unique halo twins with
$M_h{\geq}10^{14}\,\hmsol$ from the 40 paired simulations.

For the density peak associated with each twins pair, we characterise its
peak height by the halo mass $M_h$ at $z{=}0$ in the $\nuzero$ simulation
to avoid any confusion in the notations~(as $\nu$ is reserved for neutrinos
rather than peak height). The peak curvature $s$ is computed as
\begin{equation}
    s=-\frac{\Delta \,\delta_{\mathrm{cdm}}}{\Delta\, \log M},
    \label{eqn:s}
\end{equation}
where $\Delta\delta_{\mathrm{cdm}}$ is the CDM overdensity difference
between $0.75 r_{\mathrm{pk}}$ and $r_{\mathrm{pk}}$, and $\Delta\log M$ is
the logarithmic differences between the enclosed CDM masses within the two
radii.  Note that our curvature definition differs from
that of \citetalias{Dalal2008} in two aspects: 1) we compute $s$ using the
gradient of the profile of initial density peaks at $z{=}127$, instead of
the main branch of the halo merger trees used by \citetalias{Dalal2008}; 2)
Unlike \citetalias{Dalal2008}, our curvature definition has an extra minus
sign so that steeper peaks have larger positive values of $s$. As a sanity
check, Figure~\ref{fig:curvature_colormap} shows the distribution of 1000
randomly selected haloes on the $M_h$ vs. $\delta$ plane, colour-coded by
$s$ according to the colourbar on the right, where $\delta$ is the linear
theory-extrapolated Lagrangian overdensity at $z{=0}$.  Similar to the Fig.
1 in~\citetalias{Dalal2008}, while the overall distribution of $\delta$
centred around the critical overdensity $\delta_c{=}1.68$, steeper peaks
generally collapsed earlier and have higher values of $\delta$ than the
shallower ones.  To remove the average halo mass dependence of $s$ in our
analysis, we introduce the {\it relative} peak curvature $\hat{s}$ as an
HAB indicator, defined as
\begin{equation}
    \hat{s}=s-\langle s \mid M_h\rangle,
    \label{eqn:s_hat}
\end{equation}
where $\langle s | M_h\rangle$ is the average curvature as a function of
halo mass.

\begin{figure*}
	\includegraphics[width=0.9\linewidth]{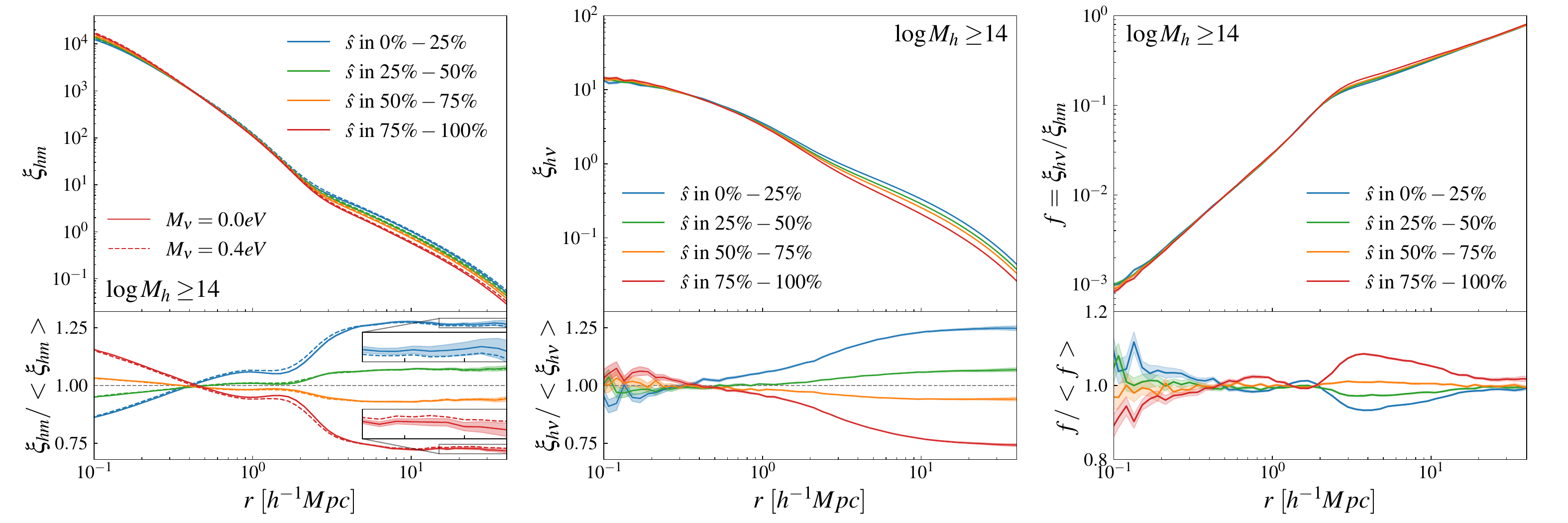}
    \caption{3D cross-correlation functions between haloes and CDM
    $\xi_{\mathrm{hm}}$~(left), haloes and neutrinos
    $\xi_{\mathrm{h\nu}}$~(middle), and the ratio between the two
    correlation functions $\xi_{\mathrm{h\nu}}/\xi_{\mathrm{hm}}$~(right),
    for haloes divided into four quartiles by their relative peak curvature
    $\hat{s}$~(listed in the top right corner of the left panel). Solid and
    dashed curves show the measurements from the $\nuzero$ and $\nufour$
    simulations, respectively. The bottom panels show the ratio between
    profiles of the individual quartiles to that of the overall
    population. The inset panels in the bottom left panel provide a zoom-in
    view of the difference between the $\nuzero$ and $\nufour$ measurements
    above 15 $\hmpc$.} \label{fig:xi}
\end{figure*}

Inspired by neutrino differential condensation discovered in
\citet{Yu2017}, we measure the neutrino-to-CDM ratio $f_{\nu}$ for each
halo in the $\nufour$ simulations.  In particular, we define $f_{\nu}$ as
\begin{equation}
    f_{\nu}\equiv\frac{\Omega_{\mathrm{cdm}}}{\Omega_\nu}\cdot
    \left[\frac{\rho_\nu(r)}{\rho_{\mathrm{cdm}}(r)}
    \right]_{r{=}5{-}12\,\mathrm{Mpc}/h},
    \label{eqn:f_nu}
\end{equation}
where $\Omega_\mathrm{cdm}$~($\Omega_\nu$) is the average CDM~(neutrino)
density of the simulation, and $\rho_{\mathrm{cdm}}$~($\rho_\nu$) is the
local CDM~(neutrino) density averaged between $r{=}5{-}12\,\hmpc$ around
each halo.  The minimum aperture of $5\,\hmpc$ is for excluding the
halo region; We have tested different maximum apertures between
$10{-}20\,\hmpc$ and verified that the impact on our results is neglible.
The radial range is set so that it is below the neutrino free-streaming
length $\lfs$ (${\simeq}20\,\hmpc$ for $\mnu{=}0.4\,\ev$) and the distance
scales that we adopt for calculating halo bias.  The average
neutrino-to-CDM ratio $\langle f_{\nu} | M_h\rangle$ deceases slightly with
increasing halo mass~(black dashed curve), with a mass-dependent scatter
$\sigma_{f_{\nu}}(M_h)$~(gray shaded band). Similar to
Equation~\ref{eqn:s_hat}, we define the neutrino excess parameter
$\epsilon_\nu$ as another indicator of the HAB, so that
\begin{equation}
    \epsilon_\nu \equiv  \frac{f_{\nu}-
    \langle f_{\nu}\mid M_h\rangle}{\sigma_{f_{\nu}}(M_h)}.
    \label{eqn:nuexcess}
\end{equation}

Our peak-based twin halo catalogue has a unique advantage for isolating the
subtle effect of massive neutrinos on HAB. Since each twins pair is tied to
the same density peak in the initial condition of both $\nuzero$ and
$\nufour$ simulations, we regard them as the {\it same} physical object
that merely exhibits different behaviors with the massive neutrinos on and
off. Therefore, if we measure two separate HAB signals using the relative
peak curvature $\hat{s}$ as the indicator in both $\nuzero$ and $\nufour$
simulations, the difference between $\hat{b}(\hat{s}|\nuzero)$ and
$\hat{b}(\hat{s}|\nufour)$ can only be induced by massive neutrinos. More
important, we can now measure $\hat{b}(\epsilon_\nu|\nuzero)$, which is the
HAB of haloes in the $\nuzero$ simulations, but using their $\epsilon_\nu$
in the $\nufour$ simulations as the HAB indicator. By comparing this HAB
signal with that measured directly from the $\nufour$ simulations
$\hat{b}(\epsilon_\nu|\nufour)$, we will be able to ascertain the {\it
direct} impact of massive neutrinos on HAB.

The left column of Figure~\ref{fig:colormap} shows the initial CDM
overdensity $\delta_{\mathrm{cdm}}^{\mathrm{ini}}$ at $z{=}127$ surrounding
two representative density peaks with equal peak height, one with a
shallow~($\hat{s}{=}{-}0.51$; top) curvature and the other
steep~($\hat{s}{=}0.40$; bottom). In each inset panel, we compare the 1D
overdensity profiles of the two peaks, with the profile of that shown in
the main panel indicated by the solid curve. Each peak subsequently
collapsed into the twin haloes in the $\nuzero$~(second column) and
$\nufour$~(third column) simulations, with their halo radii $r_{200m}$ and
characteristic scales $r_{s}$ indicated by the outer orange and inner lime
circles, respectively. The distributions of CDM particles~(black dots) are
very similar between the two simulations, and the neutrino density
$1+\delta_\nu$ roughly follows the CDM distribution on large scales in the
$\nufour$ simulation. The rightmost column shows the neutrino-to-CDM
density ratio $\rho_\nu/\rho_\mathrm{cdm}$~(normalised by
$\Omega_\nu/\Omega_{\mathrm{cdm}}$) surrounding the halos formed in the
$\nufour$ simulation.  Each panel of Figure~\ref{fig:colormap} corresponds
to a region of $30 \,\hmpc{\times}30\,\hmpc$ size centred on the peak/halo,
projected along a $12\,\hmpc$ length in the $z$-direction.  All the
distributions are colour-coded according to the respective colourbars on
top.  As expected, the steep peak collapsed into haloes with higher
concentration than the shallow one. The concentrations of haloes in the
$\nufour$ simulation is in general higher than those in the $\nuzero$
one~\citep{Brandbyge2010,Villaescusa-Navarro2013, Lazeyras2021}. This
overall concentration difference is removed from the HAB comparison by
using $\hat{c}$.  In the third column, the two haloes in the $\nufour$
simulation share very similar neutrino content {\it within} the halo
radii~(orange circles), but differ significantly in their $\delta_\nu$
distributions in the outside. However, the two $\rho_\nu/\rho_\mathrm{cdm}$
distributions shown in the fourth column are very similar on scales below
$5\,\hmpc$, while exhibiting strong discrepancies on scales above
$5\,\hmpc$.  Therefore, within the $\nufour$ simulation, the halo formed
from the high-$\hat{s}$ peak has a larger $\epsilon_\nu{=}0.57$ than that
from the low-$\hat{s}$ one~(${-}0.75$).  This signals the existence of a
positive correlation between $\hat{s}$ and $\epsilon_\nu$, hence a
potential neutrino-induced HAB.

\section{Results}
\label{sec:result}

\begin{figure*}
	\includegraphics[width=\linewidth]{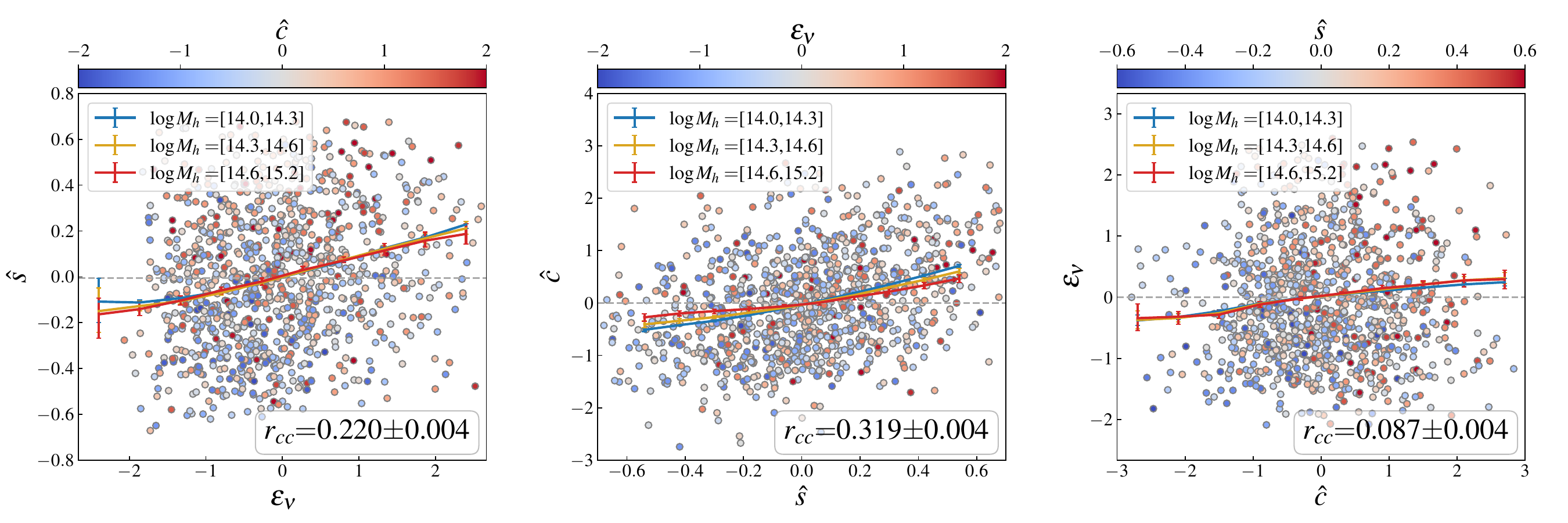}
    \caption{Pairwise correlations between $\epsilon_\nu$, $\hat{s}$, and
    $\hat{c}$. In each panel, circles represent 1000 randomly selected
    haloes from the twin halo catalogue on the axis plane defined by two
    parameters, colour-coded by their values of the third parameter. Blue,
    gold, and red curves show the mean relations between the two axis
    parameters for haloes within the three mass bins listed in the top
    right corner, while the overall Pearson's correlation coefficient
    $r_{\mathrm{cc}}$ and its $1{-}\sigma$ uncertainty are listed in bottom
    right.}
\label{fig:nuexcess_curvature_concentration}
\end{figure*}

To ascertain the correlation between $\hat{s}$ and $\epsilon_\nu$, we show
the 3D cross-correlation functions between haloes and
CDM~($\xi_{\mathrm{hm}}$; left), haloes and massive
neutrinos~($\xi_{\mathrm{h\nu}}$; middle), and the ratio between the
two~($\xi_{\mathrm{h\nu}}/\xi_{\mathrm{hm}}$; right), for haloes in
different quartiles of $\hat{s}$ in Figure~\ref{fig:xi}. In each panel, the
top main panel shows the profile comparison between four quartiles in
$\hat{s}$, while the bottom sub-panel highlights their differences using
the ratio between each quartile and the overall profile of that mass bin.
The uncertainty bands are computed by Jackknife resampling over the 40
independent simulations. Note that we have combined the results
from all halo masses above $\log M_h{>}14$ in Figure~\ref{fig:xi}, and we
have carefully checked that any trend with halo mass has been removed by
our adoption of the {\it relative} quantities.

In the left panel of Figure~\ref{fig:xi}, solid and dashed curves are the
$\xi_{\mathrm{hm}}$ measurements from the $\nuzero$ and $\nufour$
simulations, respectively.  The $\xi_{\mathrm{hm}}$ profiles exhibit the
typical one- and two-halo components that can be described using the
combination of NFW~\citep{Navarro1997} profiles and biased versions of the
matter-matter auto-correlation function
$\xi_{\mathrm{mm}}$~\citep{Hayashi2008,Zu2014}. A strong HAB effect is
present in both simulations, as shown by the bottom subpanel, where haloes
collapsed from steep peaks~(red and brown) are more concentrated within the
halo radius and have lower relative clustering amplitude on scales above
$15\,\hmpc$ than their shallow-peak counterparts~(blue and green).
However, the HAB in the $\nufour$~(dashed) simulations is slightly weaker
than that in the $\nuzero$ ones~(solid), in that the relative clustering on
scales above $15\,\hmpc$ of the steepest~(shallowest) quartile is
lower~(higher) in $\nufour$ than in $\nuzero$ simulations.  This weakening
of $\hat{s}$-indicated HAB in the $\nufour$ simulations, albeit very small
in amplitude, is robustly detected at ${\sim}1{-}\sigma$ level across all
radial bins above $15\,\hmpc$~(see the zoom-in inset panels). This
suggests that the HAB is likely modified by the presence of massive
neutrinos. The middle panel of Figure~\ref{fig:xi} is similar to the left,
but for the $\xi_{\mathrm{h\nu}}$ profiles measured from the $\nufour$
simulations.  Consistent with the results from \citet{LoVerde2014}, the
profiles of the ``neutrino haloes'' are shallower than that of the CDM
haloes. Interestingly, the $\xi_{\mathrm{h\nu}}$ profiles exhibit its own
HAB phenomenon analogous to the CDM version --- the steep density peaks
attract more concentrated ``neutrino haloes'' that exhibit lower relative
clustering of neutrinos on large scales than the shallow ones.  {\it This
to our knowledge is the first detection of the ``neutrino halo'' assembly
bias effect in simulations.}

The right panel of Figure~\ref{fig:xi} shows the ratio profiles between
$\xi_{\mathrm{h\nu}}$ and $\xi_{\mathrm{hm}}$ in the $\nufour$ simulations.  Within
$r_{200m}$, the ratio profile rises sharply as ${\simeq}r^{3/2}$, but
levels off as ${\simeq}r^{2/3}$ beyond $r_{200m}$.  Though not shown here,
the ratio profile should reach a plateau of unity
at scales ${\gg}\lfs{\simeq}20\,\hmpc$ where neutrinos follow CDM.  More important,
the bottom panel reveals that $\hat{s}$ indeed correlates with the
neutrino-to-CDM ratio on quasi-linear scales. In particular, haloes
collapsed from the steepest~(shallowest) quartile
exhibit a ${\sim}10\%$ enhancement~(deficit) in their neutrino-to-CDM ratio
between $r_{200m}$ and $\lfs$ compared to the average population.
Therefore, we expect a positive correlation between $\hat{s}$ and the
previously defined neutrino excess parameter $\epsilon_\nu$.  Given that
massive neutrinos seem to cause an impact on the $\hat{s}$-indicated
HAB~(left panel of Figure~\ref{fig:xi}), such a $\hat{s}$-$\epsilon_\nu$
correlation implies that the impact on the $\epsilon_\nu$-indicated HAB
would likely be greater.

While being theoretically informative, neither $\hat{s}$ nor $\epsilon_\nu$
is an observable quantity. As of now, concentration remains one of the most
promising observable indicators of HAB~\citep{Zu2021}, and it is thus
imperative that we explore the neutrino impact on the $\hat{c}$-indicated
HAB.  Figure~\ref{fig:nuexcess_curvature_concentration} presents the
pairwise correlations between the three HAB indicators.  For each panel of
$x$ vs. $y$, circles represent the $x$ and $y$ values of 1000 individual
haloes randomly drawn from the twin halo catalogue in the $\nufour$
simulation, each colour-coded by the value of the third indicator according
to the colourbar on top. The three coloured curves are the mean relations
$\langle y | x\rangle$ calculated for the three halo mass bins listed in
the legend, while the overall cross-correlation coefficient
$r_{\mathrm{cc}}$ is annotated in the bottom right with its $1{-}\sigma$
uncertainty.  Overall, the three indicators all correlate with one another,
with little dependence on halo mass. The correlation between $\hat{s}$ and
$\hat{c}$~(middle panel) is the strongest among the three pairs with
$r_{\mathrm{cc}}{=}0.319$. The correlation between $\hat{s}$ and
$\epsilon_\nu$ are reasonably strong with $r_{\mathrm{cc}}{=}0.220$,
consistent with our expectation from the right panel of
Figure~\ref{fig:xi}. Meanwhile, the correlation between $\epsilon_\nu$ and
$\hat{c}$ is significantly detected but with a weak strength of
$r_{\mathrm{cc}}{=}0.087{\pm}0.004$.  Although the intrinsic correlation
between $\epsilon_\nu$ and $\hat{c}$ ought to be higher, because the
measurement of $\hat{c}$ is subject to shot noise~\citep[see][for potential
improvements]{Wang2023}, the measurement error in real observations can
only be larger~\citep{Zu2017}.

\begin{figure*}
	\includegraphics[width=\linewidth]{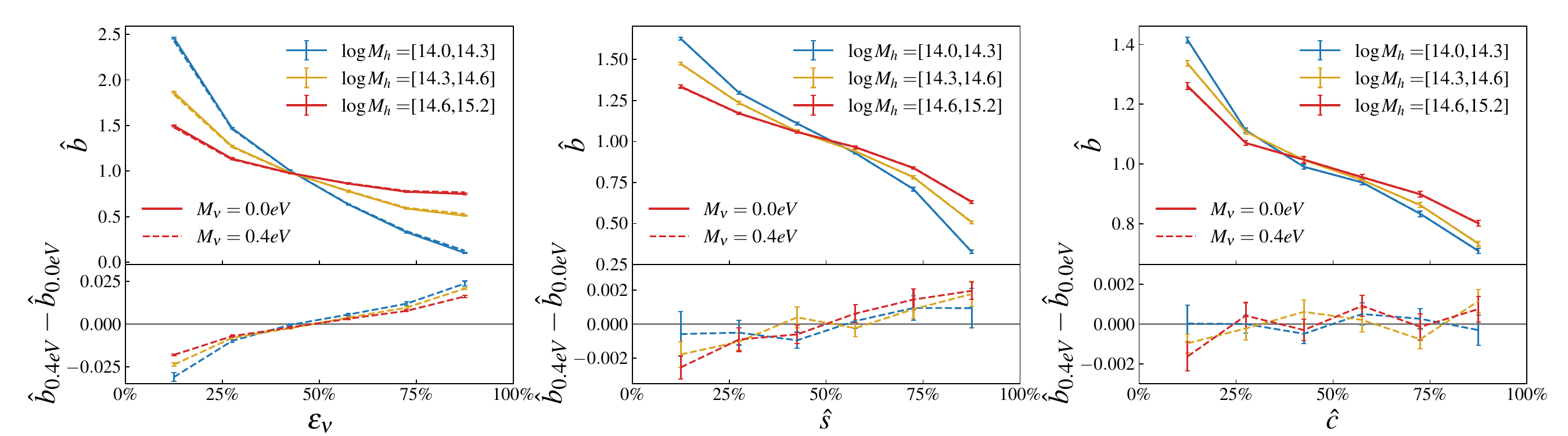}
    \caption{HAB comparison between the $\nuzero$~(solid curves) and
	$\nufour$~(dashed curve) simulations, using the neutrino excess
	$\epsilon_\nu$, relative peak curvature $\hat{s}$, relative
	concentration $\hat{c}$ as indicators in the left, middle, and
	right panel, respectively. The HAB is characterised by the relative
	bias $\hat{b}$ as a function of the indicator divided into six
	sextiles.  Blue, gold, and red curves
    indicate the results from the three halo mass bin of $\log M_h{=}[14.0,
    14.3]$, $[14.3, 14.6]$, and $[14.6, 15.2]$, respectively.  Each bottom
    subpanel shows the difference between the two sets of $\hat{b}$
    functions in the $\nuzero$ and $\nufour$ simulations. All errorbars are
    jackknife uncertainties from the 40 independent simulations. }
    \label{fig:bias}
\end{figure*}

Finally, we show the impact of massive neutrinos on the HABs indicated by
neutrino excess $\epsilon_\nu$~(left), relative peak curvature
$\hat{s}$~(middle), and relative halo concentration $\hat{c}$~(right) in
Figure~\ref{fig:bias}, the key result of this paper. In each panel of
indicator $x$~(divided into sextiles), the main panel compares the HABs as
measured by $\hat{b}(x|M_h)$ between the $\nuzero$~(solid curves) and
$\nufour$~(dashed curves) simulations, in three different mass bins of
$\log M_h{=}[14.0, 14.3]$~(blue), $[14.3, 14.6]$~(gold), and $[14.6,
15.2]$~(red).  The halo bias $b$ is measured from
$\xi_{\mathrm{hm}}/\xi_{\mathrm{mm}}$ on
scales between $15\,\hmpc$ and $30\,\hmpc$~(see the left panel of
Figure~\ref{fig:xi}), with Jackknife errorbars estimated from the 40
simulations. The discrepancy between the two sets of HABs is more clearly
seen in the bottom subpanel, which shows $\Delta \hat{b}(x|M_h) =
\hat{b}(x|M_h)_{\nufour} - \hat{b}(x|M_h)_{\nuzero}$.

When $\epsilon_\nu$ is used as the indicator~(left panel of
Figure~\ref{fig:bias}), the HAB strength is significantly weakened in the
presence of massive neutrinos. Note that even in zero-neutrino mass
simulations, we still detect an HAB signal associated with $\epsilon_\nu$.
This phenomenon can be attributed primarily to the apparent
anti-correlation between $\epsilon_\nu$ and the large-scale dark matter
overdensity. Combining the three halo mass bins, the relative bias of
the least neutrino-rich sextile of haloes is reduced by
$0.033{\pm}{0.003}$, while that of the most neutrino-rich sextile of haloes
is enhanced by $0.028{\pm}{0.001}$.  The neutrino effect decreases by an
order of magnitude when $\hat{s}$ is used as the HAB indicator~(middle
panel), in which case the $\Delta\hat{b}$ of the steepest~(shallowest)
sextile of density peaks becomes
${-}0.00105{\pm}{0.00194}$~(${}0.00207{\pm}{0.00072}$).  Unfortunately,
these $0.1{-}2$ per-cent level differences are currently only detectable in
simulations, especially given that the upper bound on $\mnu$ is well below
$0.4\,\ev$~\citep{DiValentino2021}.

When $\hat{c}$ is adopted for HAB detection~(right panel of
Figure~\ref{fig:bias}), the effect of massive neutrinos completely
diminishes, with $\Delta\hat{b}$ consistent with zero across all sextiles
of relative concentration. This is somewhat expected given that $\hat{c}$
barely correlates with $\epsilon_\nu$, which by definition is the
sole underlying driver of neutrino-induced HAB.  This null detection of
neutrino effects is also consistent with the result of
\citet{Lazeyras2021}, despite the difference in concentration measurement
and halo catalogues. Therefore, Figure~\ref{fig:bias} elucidates the impact
of massive neutrinos on concentration-indicated HAB --- massive neutrinos
are able to modify the HAB signal predicted by massless neutrino
simulations, as long as haloes are divided by their local neutrino-to-CDM
ratio or the curvature of their initial density peaks.

\section{Conclusion}
\label{sec:conclusion}

In this paper, we elucidate the impact of massive neutrinos on the halo
assembly bias~(HAB) using a large suite of massless and massive
neutrino~($\mnu{=}0.4\,\ev$) simulations from \texttt{Quijote}. By tracing
haloes back to their density peaks in the initial condition shared by both
types of simulations, we construct a twin halo catalogue where each twins
pair is characterised by their relative concentration $\hat{c}$, relative
peak curvature $\hat{s}$, and relative neutrino-to-CDM ratio
$\epsilon_\nu$.

Using the relative halo bias $\hat{b}$ to quantify the strength of HAB, we
demonstrate that the massive neutrinos could cause a two per cent-level
impact on the HAB effect compared to that predicted by the massless
neutrino simulations, if $\epsilon_\nu$ is used as the HAB indicator.
Since $\hat{s}$ is strongly correlated with $\epsilon_\nu$, the
neutrino-induced HAB remains discernible at the $0.1$-per cent level when
$\hat{s}$ is used as the indicator. However, the minuscule correlation
between $\hat{c}$ and $\epsilon_\nu$ renders the $\hat{c}$-indicated HABs
indistinguishable between the massless and massive neutrino simulations ---
a null effect consistent with previous works~\citep[e.g.,][]{Lazeyras2021}.

 An observational detection of such a massive neutrino effect on HAB is
highly unlikely within the near future. We plan to explore if there exist
halo observables that can potentially correlate better with $\epsilon_\nu$
than $\hat{c}$, thereby yielding  potentially detectable
neutrino-induced HAB signals, at least in the simulations.  Although
extremely challenging, an observational detection of the neutrino-induced
HAB with next-generation galaxy surveys, including the Dark Energy
Spectroscopic Instrument~\citep[DESI;][]{DESI2022}, Prime Focus
Spectrograph~\citep[PFS;][]{Takada2006}, Jiaotong University Spectroscopic Telescope~\citep[JUST;][]{JUST2024}, Roman Space Telescope~\citep[{\it
Roman};][]{Roman2015}, {\it EUCLID}~\citep{Euclid2011}, and the Chinese
Survey Space Telescope~\citep[{\it CSST};][]{CSST2019}, would provide an
independent path to measuring $\mnu$ using the LSS information beyond the
linear power spectrum~\citep{Bayer2022},

\section*{Acknowledgements}

We thank the anonymous referee for the many suggestions that have greatly improved the manuscript. We thank Francisco Villaescusa-Navarro, Kai Wang, Yu Liu, and Haoran Yu for
helpful discussions.  This work is supported by the National Key Basic
Research and Development Program of China (No. 2023YFA1607800,
2023YFA1607801), the National Science Foundation of China (12173024,
11890692, 11873038, 11621303), the China Manned Space Project (No.
CMS-CSST-2021-A01, CMS-CSST-2021-A02, CMS-CSST-2021-B01), and the ``111''
project of the Ministry of Education under grant No. B20019. Y.Z.
acknowledges the generous sponsorship from Yangyang Development Fund. Y.Z.
thanks Cathy Huang for her hospitality at the Zhangjiang High-tech Park.
\section*{Data Availability}

The Quijote simulation data underlying this article are publicly available
via \url{https://github.com/franciscovillaescusa/Quijote-simulations}.



\bibliographystyle{mnras}
\bibliography{ms} 





%


\bsp	
\label{lastpage}
\end{document}